\documentclass[conference]{IEEEtran}
\IEEEoverridecommandlockouts
\usepackage{pgfplots}
\usepackage{xcolor}
\usepackage{amsmath}
\pgfplotsset{compat=newest}
\usepgfplotslibrary{groupplots}
\usepackage{tikz}
\usetikzlibrary{shapes.geometric, arrows, positioning,backgrounds,fit}
\usetikzlibrary{external}
\usepgfplotslibrary{dateplot}
\usepackage{multirow}

\newcommand{\mpssq}{$\mathrm{m}/\mathrm{s}^2$}
\newcommand{\mpscu}{$\mathrm{m}/\mathrm{s}^3$}

\usepackage{cite}
\usepackage{amsmath,amssymb,amsfonts}
\usepackage{algorithmic}
\usepackage{graphicx}
\usepackage{textcomp}
\usepackage{xcolor}
\def\BibTeX{{\rm B\kern-.05em{\sc i\kern-.025em b}\kern-.08em
    T\kern-.1667em\lower.7ex\hbox{E}\kern-.125emX}}

\begin{document}

\title{Analyzing the performance of a V2X-enhanced braking system in real-world crash situations\\
}

\author{Jan Zimmermann$^{*}$, Jörg Mönnich$^{\dagger}$, Michael Scherl$^{\ddagger}$, Ignacio Llatser$^{*}$, Florian Wildschütte$^{*}$,  Frank Hofmann$^{*}$
\thanks{$^{*}$ Corporate Research - Advanced Solutions for Visual Perception, Positioning and Communication Department, Robert Bosch GmbH, 31139 Hildesheim, Germany.
{\tt\small JanChristian.Zimmermann[at]de.bosch.com}}
\thanks{$^{\dagger}$ Corporate Research - Vehicle Mechatronics Department, Robert Bosch GmbH, 70465 Stuttgart, Germany.
}
\thanks{$^{\ddagger}$ Corporate Research - Advanced Autonomous Systems Department, Robert Bosch GmbH, 70465 Stuttgart, Germany.
}

}

\maketitle

\begin{abstract}
By using an automated braking system, such as the Automatic Emergency Brake (AEB), crashes can be avoided in situations where the driver is unaware of an imminent collision. However, conventional AEB systems detect potential collision adversaries with onboard sensor systems, such as radars and cameras, that may fail in non-line-of-sight situations. By leveraging vehicle-to-everything (V2X) communication, information regarding an approaching vehicle can be received by the ego vehicle at an early point in time, even if the opponent vehicle is occluded by a view obstruction. In this work, we consider a 2-stage braking cascade, consisting of a partial brake, triggered based on V2X information, and a sensor-triggered AEB. We evaluate its crash avoidance performance in real-world crash situations extracted from the German In-Depth Accident Study (GIDAS) database using an accident simulation framework. The results are compared against a sensor-triggered AEB system and a purely V2X-triggered partial brake. To further analyze the results, we identify the crash cause for each situation in which the brake function under test could not prevent the crash. The simulation results show a high added benefit of the V2X-enhanced braking systems compared to the exclusive use of visual-based sensor systems for automated collision prevention. 
\end{abstract}

\section{Introduction}\label{sec:introduction}
Vehicle-to-everything (V2X) connects traffic participants with each other as well as with the traffic infrastructure. It has a wide field of applications in modern mobility systems like cooperative maneuvering in intersections \cite{Thunberg2021}, truck platooning \cite{Sidorenko2021} and Adaptive Cruise Control (ACC) \cite{Sidorenko2022}. Especially for active safety systems, V2X plays an important role as it can bridge a gap in state-of-the-art sensor system's information retrieval: Consider an intersection where the line of sight between two vehicles, which are approaching the intersection from different directions, is blocked by a view obstruction. Here, neither the driver nor the onboard sensor system, typically consisting of radar and video sensors, has the ability to perceive the oncoming vehicle. V2X communication, however, is not bound to an unobstructed view and, thus, information can be provided to the ego vehicle, which can be used to trigger an automated braking system and prevent a possible crash between the vehicles.

As the Automatic Emergency Brake (AEB) is a full force brake, the requirements regarding functional safety are high, because a wrongful execution of this brake can create dangerous situations. Therefore, in order to activate the AEB, the information, based on which the brake is triggered, needs to fulfill specific functional safety requirements, named Automotive Safety Integrity Levels (ASIL). To the current state of the art, standardized V2X messages like the cooperative awareness message (CAM) or cooperative perception message (CPM) do not incorporate any safety aspects \cite{Llatser2023}, which means that received V2X data cannot be used to trigger an AEB. This aspect needs to be considered when designing a serial V2X-enhanced braking system. 

Although the benefit of additional sensing capabilities in the form of V2X communication is beyond dispute, a quantification of the benefit is an important factor that can drive the development of such systems. Furthermore, the limitations of V2X-enhanced braking systems need to be analyzed to further improve the system.

\subsection{State of the Art}\label{subsec:stateoftheart}
There already exists a broad body of research dedicated to the benefit analysis of onboard sensor triggered AEB systems \cite{Sander2019} \cite{Scanlon2017}. Regarding the use of V2X information for automatic (emergency) braking systems, the research so far was mainly focused on rear-end collision avoidance \cite{Sidorenko2021} \cite{Zhang2021} \cite{Pimentel2023}, while less work has been dedicated to intersections. However, about 20~\% of fatal accidents occur in intersections in the European Union and the United States \cite{Khayyat2022}. 

The works \cite{Khayyat2022} and \cite{Lai2023} propose a V2X-based AEB system for intersections. In the respective papers, the brake force is adjusted, varying between a partial brake and a full force brake, depending on distance or time of the vehicle to the estimated crash. While in \cite{Khayyat2022} an onboard radar system is assumed to be installed in the ego vehicle next to communication capabilities, the system described in \cite{Lai2023} relies on V2X information alone. The above mentioned papers assume that a full force brake can be activated based on information received via communication\footnote{In \cite{Khayyat2022}, the information received via the onboard sensor system and V2X are fused together, making it impossible to distinguish the different information sources.}. However, under the current state of the art, this is not always possible due to functional safety regulations. The authors of \cite{Avino2018} analyze the crash avoidance potential by means of V2X-based driver warnings, with a focus on the effect of the V2X penetration rate as well as on imperfect channel effects.

In an earlier work \cite{Zimmermann2024}, we proposed a 2-stage braking system that takes the safety requirements of the different information sources into account. The crash avoidance and mitigation capabilities are demonstrated by simulation in Euro NCAP related 90° crossing situations, while constant velocity of the approaching vehicles is assumed. This is also the case for the already mentioned work in  \cite{Khayyat2022} \cite{Lai2023}.
However, a broad investigation of crash avoidance capabilities in diverse and realistic situations (regarding intersection, opponent and ego behavior), such as it was performed for example for the onboard sensor triggered AEB in \cite{Scanlon2017}, is currently still missing for V2X-enhanced braking systems. 
\subsection{Contributions}\label{subsec:contributions}
Considering the state of the art described above, we make the following contributions within this work:
\begin{enumerate}
    \item We evaluate and compare the crash avoidance performance of an onboard sensor-triggered AEB, a safety-conform V2X partial brake and a 2-stage braking system\footnote{We introduced this 2-stage braking system in our earlier publication \cite{Zimmermann2024}.}, which comprises the two previous braking types. In the analysis, we test three different onboard sensor sets and various parameterizations regarding the earliest possible activation time point for the V2X partial brake. 
    \item As test cases, we consider real-world crash situations extracted from the German in-depth accident study (GIDAS) \cite{GIDAS} database. This approach allows us to account for real-world effects, such as reduced road friction and realistic opponent behavior in the evaluation of crash avoidance capabilities. This provides a higher level of realism regarding crash avoidance capabilities in intersections compared to Euro NCAP-related test cases, as evaluated in previous works  \cite{Khayyat2022} \cite{Lai2023} \cite{Avino2018} \cite{Zimmermann2024}.
    \item  Additionally, we identify the most probable crash cause for the crashes that could not be avoided by the brakes under test. To achieve this, we define seven different crash causes that can be evaluated automatically for every simulated case. This provides insights for future optimization of the brakes under test.
\end{enumerate}
The remainder of this paper is structured as follows. In the next Section~\ref{sec:v2x-enhanced_braking_system}, we introduce the V2X-enhanced braking system, detailing the assumptions regarding the vehicles, the braking scheme, and the triggering conditions. In the subsequent Section~\ref{sec:evaluation_procedure}, we present the evaluation procedure, including an overview of the test case set, the simulation environment, and the crash causes. In Section~\ref{sec:results}, we explain the evaluation results in detail, before concluding the paper in Section~\ref{sec:conclusion}.

\section{V2X-enhanced braking system}\label{sec:v2x-enhanced_braking_system}

\subsection{Vehicles, equipment and V2X communication}\label{subsec:vehicles_equipment_and_v2x_communication}
The focus of this analysis lies on crash situations involving two vehicles: an ego vehicle that carries the brake system under test, and an opponent vehicle which serves as the adversary of the ego vehicle in a crash scenario. 
The ego vehicle is assumed to be a passenger car and equipped with an onboard sensor system. We consider three different classes of sensor sets: a minimum class set, consisting of a single video sensor (1V) , a medium class set, consisting of one radar and one video sensor (1R~/~1V) and a premium class set, consisting of five radars and one video sensor (5R~/~1V). 
The opponent vehicle can either be a passenger car or an (electric) bicycle. In contrast to the ego vehicle, it is not assumed that the opponent vehicle has onboard radar and/or video sensing capabilities, nor does it necessarily carry an active safety system. 
Both ego and opponent vehicles are equipped with communication systems, allowing them to exchange information via V2X with each other or other traffic participants on the road\footnote{If the opponent is not equipped with a communication unit, it can possibly still be detected by an infrastructure-based sensor system. The extracted information about the opponent can then be provided by the infrastructure system over V2X to the ego vehicle.}, effectively assuming a V2X penetration rate of 100~\%. The exchanged messages are assumed to be the Cooperative Awareness Message (CAM), standardized in Europe, or the the Basic Safety Message (BSM), standardized in the US, and carry vehicle state information such as type, position, velocity, acceleration and heading. For an overview over the standardized V2X messages and assumed V2X technology, we refer to our earlier publication \cite{Zimmermann2024}. Additionally, we assume a perfect communication channel between the vehicles and therefore neglect effects such as shadowing and fast fading.

%

\subsection{Braking scheme}\label{subsec:braking_scheme}
The V2X-enhanced braking system is structured as a 2-stage braking cascade. The first stage is a partial brake, i.e., a brake with a reduced braking force of 4~\mpssq, triggered by information received via V2X or possibly from the onboard sensor system. A brake force up to this level is considered not to require an ASIL classification. The second stage is an onboard sensor-triggered automatic emergency brake (AEB), which is a full-force brake that decelerates the vehicle at 9~\mpssq\ and, therefore, requires an ASIL classification.
\begin{figure}
    \centering
    \scalebox{.8}{\begin{tikzpicture}

\definecolor{firebrick1702141}{RGB}{214,39,40}
\definecolor{darkorange}{RGB}{255,127,14}
\definecolor{steelblue}{RGB}{30,96,164}
\definecolor{lightseagreen64183173}{HTML}{2CA02C}
\definecolor{purple}{HTML}{9467BD}
\definecolor{firebrick1702141}{RGB}{170,21,41}
\definecolor{lightseagreen6418317364183173}{RGB}{64,183,173}

\tikzstyle{my below of} = [below=of #1.south]
\tikzstyle{dist below of} = [below=of #1.south, node distance = 1cm]
\tikzstyle{my right of} = [right=of #1.east]
\tikzstyle{my left of} = [left=of #1.west]
\tikzstyle{my above of} = [above=of #1.north]

\tikzstyle{arrow} = [thick,->,>=stealth]

\tikzstyle{interface} = [rectangle, rounded corners, minimum width=1cm, minimum height=0.4cm,text centered, draw=black, fill=firebrick1702141!30, font=\scriptsize]
\tikzstyle{function} = [rectangle, minimum width=3.5cm, minimum height=0.75cm,text centered, draw=black, fill=lightseagreen64183173!50, font=\scriptsize]
\tikzstyle{generic} = [rectangle, minimum width=3.5cm, minimum height=0.75cm,text centered, draw=black, fill=purple!30, font=\scriptsize]
\tikzstyle{AEB} = [rectangle, minimum width=3.5cm, minimum height=4.5cm,text centered, draw=black, fill=steelblue!30, font=\scriptsize]
\tikzstyle{textblock} = [rectangle,text centered, font=\scriptsize]

\node (v2xobjects) [interface] {V2X objects};
\node (egostate) [interface, right of=v2xobjects, node distance=2.5cm] {Ego state};
\node (sensorobjects) [interface, right of=egostate, node distance=2.5cm] {Sensor objects};


\node (1ststage) [textblock,  dist below of=v2xobjects] {1st stage \    \ \ \ \ \ \ \ \ \ \    partial brake};
\node (fusion) [function,  below of=1ststage, align=center, node distance = 0.75cm] {Fusion/\\ scene understanding};
\node (collisionpred) [function,  below of=fusion, node distance = 1cm] {Collision prediction};
\node (brakingdecision) [function,  below of=collisionpred, node distance = 1cm] {Braking decision};
\node (driverveto) [function,  below of=brakingdecision, node distance = 1cm] {Driver override};
\scoped[on background layer]{\node [fit=(1ststage)(driverveto), fill=lightseagreen64183173!30, draw=black] {2nd stage};}

\node (aeb) [AEB, dist below of=sensorobjects, align=center, node distance = 2cm] {2nd stage: AEB};

\node (interventionman) [generic, below  of=egostate, node distance = 7cm] {Intervention manager};

\draw [arrow] node{} (-0.25cm,-0.22)  -- node{} (-0.25cm,-1.85cm);

\draw [arrow]  (egostate.south) -- node{} (2.5,-0.5cm) -- node{} (4.75cm,-0.5cm) -- node{} (4.75cm,-1.25cm) ;
\draw [arrow] (egostate.south) -- node{} (2.5cm,-0.5cm) -- node{} (0cm,-0.5cm) -- node{} (0cm,-1.85cm);

\draw [arrow] (sensorobjects.south) -- (aeb.north);
\draw [arrow] (sensorobjects.south) -- node{} (5,-0.75cm) -- node{} (0.25,-0.75cm) -- node{} (0.25cm,-1.85cm);

\draw [arrow] (fusion.south) -- (collisionpred.north);
\draw [arrow] (collisionpred) -- (brakingdecision);
\draw [arrow] (brakingdecision) -- (driverveto);

\draw [arrow] (aeb.south) -- node{} (5,-6cm) -- node{} (3,-6cm) -- node{} (3,-6.64cm);
\draw [arrow] (driverveto.south) -- node{} (0,-6cm) -- node{} (2,-6cm) -- node{} (2,-6.64cm);

\end{tikzpicture}}
    \caption{Decision process for 2-stage brake in the ego vehicle.}
    \label{fig:flow_diagram_twostage_brake}
\end{figure}
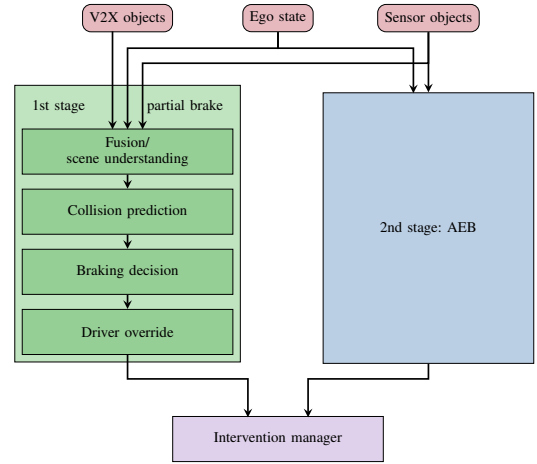

In Figure~\ref{fig:flow_diagram_twostage_brake}, the decision process regarding the two stages are depicted as a flow diagram. For the partial brake, the information from the objects, detected via V2X and onboard sensors, as well as ego state information is fused into a common scene understanding. Based on this, the ego vehicle predicts if its own trajectory intersects the estimated opponent trajectory. If a crash is predicted, the brake triggering conditions, described in the subsequent Subsection~\ref{subsec:brake_trigger_conditions}, are evaluated and it is decided whether the brake should be activated. Before the decision is passed on to the intervention manager that activates the braking system, the driver has the possibility to overwrite the braking decision of the system, e.g. by braking harder. The decision process regarding the AEB is similar to the partial brake, except that it only uses sensor information.

The relevant parameterization of the brakes is summarized in Table~\ref{tab:brake_paremeters}. The detection delay accounts for the time that the respective sensor system needs to classify an opponent vehicle as a relevant object. For the V2X system, this includes the message generation, communication latency, message reception, and information processing. The brake application delay accounts for the reaction time between the decision to trigger a brake and the actual application of the brake force.  The time-to-collision (TTC) threshold restricts the time window before the estimated crash in which the respective brake is allowed to be activated. The threshold for the ego acceleration restricts the brake application such that the brake is not activated if the ego vehicle accelerates beyond the threshold value. The safety distance characterizes the minimum distance between the ego and opponent vehicle paths at which the ego vehicle should stop under successful brake activation.

\begin{table}[]
    \centering
    \caption{Brake parametrization}
    \begin{tabular}{l|l l}
         & V2X partial brake   & AEB \\ \hline
      Max. deceleration   & 4~\mpssq & 9~\mpssq \\
      Jerk & 45~\mpscu & 45~\mpscu \\
      Brake appl. delay & 120 ms & 120 ms \\
      TTC threshold & var. 2~s - 1.25~s& 1.25~s \\
      Ego accel. threshold & 1~\mpssq&1~\mpssq \\
      Safety dist. & 0.5~m  & 0.5~m
    \end{tabular}
    
    \label{tab:brake_paremeters}
\end{table}


\subsection{Brake trigger conditions}\label{subsec:brake_trigger_conditions}
As a general restriction, the V2X partial brake and the AEB are only activated if the ego vehicle predicts a front crash with the opponent and if the driver does not react to the situation in due time. Given this restriction and focusing on intersections, the brake activation is determined such that the vehicle is stopped before the ego vehicle enters the tube enclosing the opponent vehicle and its trajectory, if possible.
Regarding the \textbf{V2X partial brake}, the following conditions must be met for brake activation: 
\begin{itemize} 
    \item  \textit{Detected:} The opponent is detected by the ego vehicle via V2X, i.e., the ego vehicle receives messages from the opponent vehicle declaring its existence and state.
    \item  \textit{Crash predicted:} Provided that an opponent is detected, a crash between the vehicles is predicted to occur during the prediction period. 
    \item  \textit{Below TTC threshold:} The time-to-collision (TTC) threshold restricts the allowed brake activation time window before the crash to a certain span, i.e., the brake can only be activated if the current TTC is below the threshold. 
    \item  \textit{TTB $\leq$ 0:} To reduce false-positive brake activations, the triggering time point of the brake is delayed until the last possible moment at which the crash can still be avoided by activating the brake. This time point is labeled as the time-to-brake (TTB) threshold and can be calculated by comparing the estimated distance to crash  $x_\mathrm{crash}$ to the required stopping distance $x_\mathrm{stop}$ of the ego vehicle. The brake needs to be activated when $x_\mathrm{crash} = x_\mathrm{stop}$ in order to prevent the crash. 
    \item  \textit{No ego acceleration: } The brake is only allowed to be triggered if the ego vehicle’s acceleration is below a certain threshold. 
\end{itemize}
The \textbf{AEB triggering conditions} regarding \textit{Crash predicted}, \textit{Below TTC threshold}, \textit{TTB $\leq$ 0} and \textit{No ego acceleration} are the same as for the V2X partial brake, although the parameterization may vary. The following conditions complement the AEB condition list: 
\begin{itemize} 
    \item  \textit{Detected:} The opponent is detected by the ego vehicle via the onboard sensor system, i.e., the radar and/or video sensor detects the opponent as a relevant object and provide information about its state.
    \item \textit{No time-to-evade:}  To reduce false-positive brake activations, the activation of the braking system is delayed until the driver cannot avoid the crash by an evasive steering maneuver, i.e., a predefined lateral acceleration.
\end{itemize}

\section{Evaluation procedure}\label{sec:evaluation_procedure}

\subsection{GIDAS and PCM test data set}\label{subsec:GIDAS_and_PCM_data}
The German In-Depth Accident Study (GIDAS) \cite{GIDAS}  is a project run by Bundesanstalt für Straßenwesen (BASt) and Forschungsvereinigung Automobiltechnik e.V. (FAT). The overall goal of this project is to improve road safety in Germany. To achieve this, a database is maintained in which crashes with personal injury, that occurred in the recording areas of Hannover and Dresden, are documented. Overall, more than 2~500 parameters are recorded per crash, including an in-depth description of the crash situation and vehicles, all persons involved in the crash, their level of injury, and the trajectories of the vehicles involved.  
Many of the GIDAS crashes are available in the so-called pre-crash matrix (PCM) format \cite{PCM}, which can be loaded into a simulation environment by a suitable interface.
\begin{figure}[h]
    \centering
    \includegraphics[width=.85\linewidth]{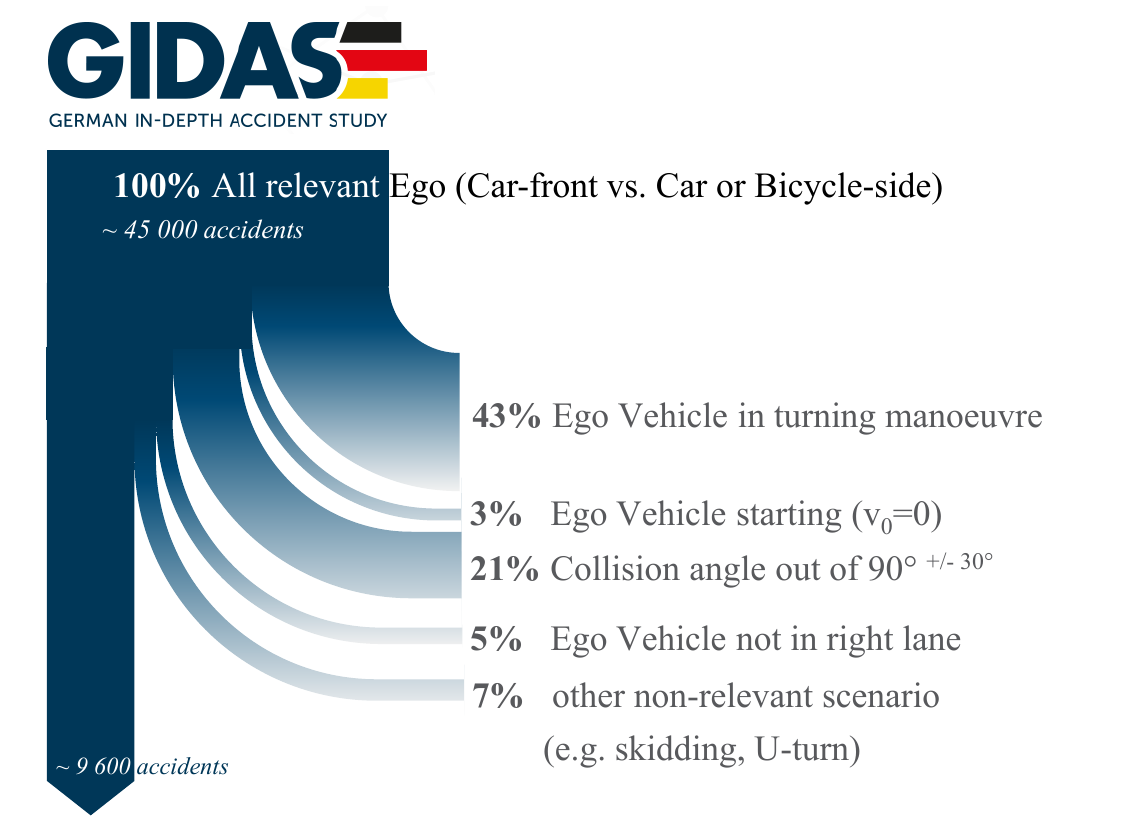}
    \caption{Applied filters to the GIDAS database. The resulting 9600 accidents correspond to 925 available PCM cases.}
    \label{fig:gidas_filter}
\end{figure}

To extract the GIDAS cases that correlate with the field of effect of intersection automatic braking systems, the data base is filtered according to Figure~\ref{fig:gidas_filter}. 
Firstly, only front-to-side crashes are considered. Additionally, the ego vehicle type is restricted to a passenger car, while only passenger cars and bicycles are admitted as opponent types. These resulting cases represent 100~\% of relevant crashes and are equal to about 45~000 crash occurrences with personal injury in Germany per year, which accounts for about 16~\% of the total annual frequency of accidents involving personal injury in Germany. 
As indicated in Figure~\ref{fig:gidas_filter}, further filters are applied. The remaining cases correspond to 9~600 crash occurrences with personal injury in Germany per year, which represents 3~\% of the overall crash occurrences with personal injury in Germany per year. Representative of this proportion, 925 cases are available in PCM format and can therefore be simulated without further transcription.

\subsection{Simulation environment}\label{subsec:simulation_environment}
\begin{figure}[h]
    \centering
    \includegraphics[width=0.5\linewidth]{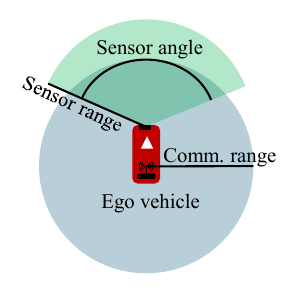}
    \caption{Ego vehicle with conic field of view of onboard sensor system (green) and circular sensor field of view of V2X 'sensor' (blue).}
    \label{fig:sensor_systems_ego}
\end{figure}
The python-based simulator employed for this study contains a module that can automatically convert PCM cases into the encoded traffic scenarios, including detailed specifications regarding the involved vehicles, their trajectories, possible view obstructions as well as road conditions. The AEB, V2X-based partial brake and the 2-stage brake function are implemented in the ego vehicle model. The situation is simulated with a step size of 10 ms, where in every time step the brake triggering conditions, see Subsection~\ref{subsec:brake_trigger_conditions}, are evaluated. If all conditions are fulfilled, the respective brake is activated and the longitudinal velocity of the ego vehicle deviates from the originally recorded velocity accordingly, i.e. the simulation is closed-loop. The vehicle's dynamics are abstracted to single track models, while the onboard sensor's field of view is modeled as a sensor cone, characterized by a sensor angle and range, see Figure~\ref{fig:sensor_systems_ego}.  Detection, classification, and sensor fusion are accounted for by a time delay, denoted as $\delta_t$. Furthermore, the V2X information transmission process is modeled as an extra sensor on the ego vehicle, featuring a 360° field of view and a sensor range equivalent to the expected V2X communication range, as indicated in Figure~\ref{fig:sensor_systems_ego}. As soon as the opponent vehicle enters the sensor field of view of the V2X sensor, its state becomes known to the ego vehicle after a detection delay $\delta_t$, which accounts for the time until the first V2X message is received and processed by the ego vehicle. This state of the opponent is available to the ego vehicle until the opponent vehicle leaves the sensor field of view again. 

The parameterization of both the V2X sensor and the onboard sensor systems is detailed in Table~\ref{tab:parameterization_sensor_systems}. The communication range of 56 m is the euclidean distance between the vehicles and results from a literature research regarding communication in intersections. At this distance, the packet delivery ratio was above 90~\%. The range of the onboard sensor systems is chosen as in \cite{Scanlon2017}\footnote{We also simulated a sensor range of 50 m. No significant difference showed compared to the results using a 120 m range.}, while the opening angles and detection delays $\delta_t$ are the result of expert discussions.

\begin{table}[]
\caption{Parameterization of sensor sets. The mount point is specified in meters from the front of the vehicle, located at the center of the vehicle's width. *For bicycles the antenna is assumed to be at $1/2$ of the bicycle’s length.}
    \centering
    \begin{tabular}{l|l l l l l}
               & Angle & Range & Mount point              & Point of recogn. & $\delta_t$  \\ \hline
       V2X     & 360°  & 56 m  & 3/4 opp. len. & 3/4 opp. length* & 300 ms \\
       1 V     & 100°  & 120 m  & 1.40 m                   & 1/2 opp. length & 200 ms\\
       1R / 1V & 120°  & 120 m  & 0.25 m                   & 1/2 opp. length &  200 ms\\
       5R / 1V & 240°  & 120 m  & 0.25 m                   & 1/2 opp. length & 200 ms
    \end{tabular}
    
    \label{tab:parameterization_sensor_systems}
\end{table}

\subsection{Crash cause analysis}\label{subsec:crash_cause_analysis}
Regarding the extracted situations from the GIDAS data base, we define the crash cause as the reason why the automatic brake under test fails to avoid the crash in simulations. Note that the crash cause does specifically not describe the original cause of the crash in the real world. 

The first considered crash causes are directly related to the triggering conditions of the brake, which are described in Subsection~\ref{subsec:brake_trigger_conditions}. The most decisive time point of the activation process is the Time-to-Brake (TTB) threshold, i.e., the time point at which the brake needs to be activated to just avoid the crash by a minimal safety distance. If the brake activation is delayed beyond this point, the distance to the crash becomes smaller than the minimal stopping distance. Therefore, assuming no other influences like hard braking of the opponent and decisive steering, the crash is physically bound to happen. With this, it can be stated that if the activation of the braking system is delayed beyond the TTB threshold for some reason, then this reason is highly likely to be the cause for the crash. As all causes that fall into this category are closely connected to the crash triggering conditions, we label these as \textbf{trigger-related causes}. The following list summarizes the trigger-related crash causes examined in this analysis. 
\begin{itemize}
    \item \textit{Detection:} Brake activation is delayed beyond the theoretical TTB threshold because the opponent is not detected by the ego vehicle in time.  Late detection can happen due to a view obstruction and limited sensor field of view or limited communication range, respectively. 
    \item \textit{Time-To-Evade (TTE) condition:} Brake activation is delayed beyond TTB threshold because the crash could still be avoided by a predefined lateral acceleration, i.e., the driver could still avoid the crash by evasive steering. This cause is only relevant if the TTE condition is an active trigger condition of the brake. 
    \item \textit{Time-To-Crash (TTC) threshold:} The TTC threshold limits the allowed brake application time window in such a way that the TTB threshold does not fall inside the window, i.e. the beginning of the window lies beyond the TTB threshold. 
    \item \textit{Ego acceleration:} The brake is delayed beyond the TTB threshold (or even not activated at all) because the ego vehicle is accelerating above a certain limit. 
\end{itemize}
Regarding these trigger-related crash causes, only the most critical is evaluated, which is defined as the one that has the highest temporal distance to the TTB threshold, i.e. delays the brake activation the furthest. Therefore, it is not possible that more than one trigger-related cause is identified.

The subsequently defined friction crash cause can occur if the friction coefficient of the road, denoted as $\mu$, is below 1. If the ego vehicle is not aware of the reduced friction, the TTB threshold is incorrectly calculated.
\begin{itemize}
    \item \textit{Friction: } Due to low and unknown surface friction, the requested brake force cannot be applied, leading to higher stopping distances than estimated and therefore to a crash. 
\end{itemize}
The calculation of the TTC as well as the TTB threshold is dependent on the estimations of the crash time point and position. However, the vehicles might not behave as predicted: Velocity can change, steering may occur, etc. Because of these types of actions, the actual crash position might deviate from the estimation, which would, for example, require an earlier start of the brake. 
\begin{itemize}
    \item \textit{Steering: } Either the opponent's or ego vehicle’s heading changes significantly within the allowed brake activation window restricted by the TTC threshold. This can lead to a late crash prediction as well as a wrong estimation of the crash time point and position.
    \item \textit{Opponent acceleration\footnote{Acceleration of the ego vehicle is accounted for by the trigger-related cause \textit{Ego acceleration}.}:} Opponent accelerates significantly in a time span, in which the opponent's state is known to the ego vehicle. This can lead to late crash prediction and wrong estimation of crash time point and position. 
\end{itemize}
For the analysis of these trigger-related causes, specific conditions are formulated that can be automatically evaluated. If such a condition is met, it is highly likely that the associated crash cause, as defined above, is the reason for the crash. 
\section{Results of performance analysis}\label{sec:results}
Three different brake types are tested in  simulation and compared against each other: An onboard sensor-triggered AEB, a V2X-triggered partial brake, and the combination of these two brake types in a 2-stage brake. Additionally, the onboard sensor set is varied between 1V, 1R~/~1V and a 5R~/~1V configuration. The resulting sensor angles and ranges associated with these sets are summarized in Table~\ref{tab:parameterization_sensor_systems}, together with their mount points on the car as well as the point of recognition of the opponent. The TTC threshold of the V2X brake, described in Subsection~\ref{subsec:braking_scheme}, is varied between 2~s, 1.5~s and 1.25~s. The brake parameters can be found in Table~\ref{tab:brake_paremeters}.
Before describing the simulation outcomes in the following sections, it is important to clarify that all results are relative to the test set of 925 PCM cases, extracted from the GIDAS database, described in the Subsection~\ref{subsec:GIDAS_and_PCM_data}.

\subsection{Overall crash avoidance performance}\label{subsec:overall_crash_avoidance}

%

\begin{table}[]
    \centering
    \caption{Overall results for crash avoidance performance. Results are represented as avoided crashes in percent of the 925 simulated PCM cases. Parameter variation includes sensor sets (1V, 1R~/~1V, 5R~/~1V) and TTC thresholds for the V2X brake (2~s, 1.5~s, 1.25~s). Road friction is unknown to braking system.}
    \begin{tabular}{c|c | c  c  c |  c c  c}
                 & AEB       & \multicolumn{3}{c|}{V2X partial brake}  & \multicolumn{3}{c}{2-stage brake}  \\ 
                 &           & 2 s         & 1.5 s  & 1.25 s   & 2 s     &1.5 s     & 1.25 s   \\ \hline 
      1 V        & 33.9\%    & 74.4\%      & 57.1\% & 41.0\%   & 88.2\%  & 81.7\%   & 75.2\%  \\
     1 R / 1 V   & 36.4\%    & 74.4\%      & 57.1\% & 41.0\%   & 88.3\%  & 81.9\%   & 75.5\% \\
     5 R / 1 V   & 38.9\%    & 74.4\%      & 57.1\%     & 41.0\%       & 88.4\%  & 82.1\% & 75.6\%  \\
    \end{tabular}
    
    \label{tab:overall_results}
\end{table}
In Table~\ref{tab:overall_results}, the overall crash avoidance performance is summarized. This performance is measured as avoided crashes in percent, relative to the 925 PCM cases that mark 100~\%, and do not indicate the overall crash avoidance potential in Germany. The crash avoidance performance for the AEB ranges between 33.9~\% and 38.9~\%, depending on the available sensor set. Therefore, with five additional radars to the video sensor, widening the field of view from 120° to 240°, about 5~\% of crashes can be avoided additionally regarding the test data set.\\
The performance of the V2X brake is independent of the sensor set. However, a strong dependence on the activation threshold, i.e., the TTC threshold, is observed. For this brake type, the performance ranges between 41.0~\% and 74.4~\%, which is a significant difference of 33.4~\%.\\
The 2-stage brake performance is dependent on both the available sensor set as well as the allowed activation threshold of the 2-stage brake. However, it can be observed that the dependence of the results on the V2X TTC threshold is significantly higher than the dependence on the available sensor set: While the performance difference between the different sensor sets is at most 0.4~\%, the performance difference regarding the different TTC thresholds can reach up to 13.0~\%. Apparently, restricting the activation windows from 2~s to 1.25~s imposes a high constraint on the performance, while decreasing the field of view does not.
Overall, the AEB is outperformed by both the 2-stage brake and the V2X brake alone in the tested scenarios. The crash avoidance performance difference can reach levels as high as 54.3~\%.

\subsection{Results for crash cause analysis}\label{subsec:results_crash_cause_analysis}
In this section, the remaining crashes are categorized according to their most probable crash causes, which are defined in Subsection~\ref{subsec:crash_cause_analysis}. The results of this analysis are represented as bar plots, see for example Figure~\ref{fig:AEB_crash_cause}:  The separate colored bars show the percentage of crash causes that fall into the respective category or, in the case of the green bar on the right side of the plot, the percentage of avoided crashes.

\subsubsection{AEB}\label{subsubsec:AEB}
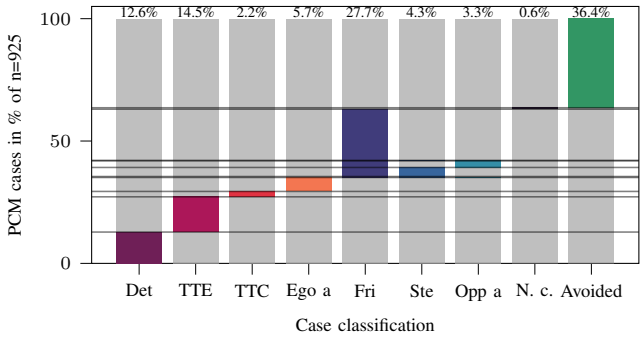
\begin{figure}[h]
    \centering
\begin{tikzpicture}

\definecolor{brown1722389}{RGB}{172,23,89}
\definecolor{coral24311881}{RGB}{243,118,81}
\definecolor{crimson2255066}{RGB}{225,50,66}
\definecolor{darkgray176}{RGB}{176,176,176}
\definecolor{darkslateblue6460123}{RGB}{64,60,123}
\definecolor{darkslategray463058}{RGB}{46,30,58}
\definecolor{gray}{RGB}{128,128,128}
\definecolor{gray127}{RGB}{127,127,127}
\definecolor{mediumaquamarine139217178}{RGB}{50,150,100}
\definecolor{purple1113187}{RGB}{111,31,87}
\definecolor{steelblue51142167}{RGB}{51,142,167}
\definecolor{steelblue55101157}{RGB}{55,101,157}

\begin{axis}[
tick align=outside,
tick pos=left,
title={},
x grid style={darkgray176},
xlabel={Case classification},
xmin=-0.84, xmax=8.84,
xtick style={color=black},
xtick={0,1,2,3,4,5,6,7,8},
xticklabels={
  Det,
  TTE,
  TTC,
  Ego a,
  Fri,
  Ste,
  Opp a,
  N. c.,
  Avoided
},
y grid style={darkgray176},
ylabel={PCM cases in \% of n=925},
ymin=0, ymax=105,
ytick style={color=black}, 
width=\linewidth,
height=5cm,
font=\scriptsize
]
\draw[draw=none,fill=gray127] (axis cs:-0.4,0) rectangle (axis cs:0.4,0);
\draw[draw=none,fill=gray127] (axis cs:0.6,0) rectangle (axis cs:1.4,0);
\draw[draw=none,fill=gray127] (axis cs:1.6,0) rectangle (axis cs:2.4,0);
\draw[draw=none,fill=gray127] (axis cs:2.6,0) rectangle (axis cs:3.4,0);
\draw[draw=none,fill=gray127] (axis cs:3.6,0) rectangle (axis cs:4.4,0);
\draw[draw=none,fill=gray127] (axis cs:4.6,0) rectangle (axis cs:5.4,0);
\draw[draw=none,fill=gray127] (axis cs:5.6,0) rectangle (axis cs:6.4,0);
\draw[draw=none,fill=gray,fill opacity=0.5] (axis cs:-0.4,0) rectangle (axis cs:0.4,100);
\draw[draw=none,fill=gray,fill opacity=0.5] (axis cs:0.6,0) rectangle (axis cs:1.4,100);
\draw[draw=none,fill=gray,fill opacity=0.5] (axis cs:1.6,0) rectangle (axis cs:2.4,100);
\draw[draw=none,fill=gray,fill opacity=0.5] (axis cs:2.6,0) rectangle (axis cs:3.4,100);
\draw[draw=none,fill=gray,fill opacity=0.5] (axis cs:3.6,0) rectangle (axis cs:4.4,100);
\draw[draw=none,fill=gray,fill opacity=0.5] (axis cs:4.6,0) rectangle (axis cs:5.4,100);
\draw[draw=none,fill=gray,fill opacity=0.5] (axis cs:5.6,0) rectangle (axis cs:6.4,100);
\draw[draw=none,fill=gray,fill opacity=0.5] (axis cs:6.6,0) rectangle (axis cs:7.4,100);
\draw[draw=none,fill=gray,fill opacity=0.5] (axis cs:7.6,0) rectangle (axis cs:8.4,100);
\draw[draw=none,fill=purple1113187] (axis cs:-0.4,0) rectangle (axis cs:0.4,12.8);
\draw[draw=none,fill=brown1722389] (axis cs:0.6,12.8) rectangle (axis cs:1.4,27.2);
\draw[draw=none,fill=crimson2255066] (axis cs:1.6,27.2) rectangle (axis cs:2.4,29.4);
\draw[draw=none,fill=coral24311881] (axis cs:2.6,29.4) rectangle (axis cs:3.4,35.1);
\draw[draw=none,fill=darkslateblue6460123] (axis cs:3.6,35.1) rectangle (axis cs:4.4,35.5);
\draw[draw=none,fill=steelblue55101157] (axis cs:4.6,35.1) rectangle (axis cs:5.4,35.5);
\draw[draw=none,fill=steelblue51142167] (axis cs:5.6,35.1) rectangle (axis cs:6.4,35.5);
\draw[draw=none,fill=darkslateblue6460123] (axis cs:3.6,35.5) rectangle (axis cs:4.4,39.2);
\draw[draw=none,fill=steelblue55101157] (axis cs:4.6,35.5) rectangle (axis cs:5.4,39.2);
\draw[draw=none,fill=darkslateblue6460123] (axis cs:3.6,39.2) rectangle (axis cs:4.4,41.9);
\draw[draw=none,fill=steelblue51142167] (axis cs:5.6,39.2) rectangle (axis cs:6.4,41.9);
\draw[draw=none,fill=steelblue55101157] (axis cs:4.6,41.9) rectangle (axis cs:5.4,42.1);
\draw[draw=none,fill=steelblue51142167] (axis cs:5.6,41.9) rectangle (axis cs:6.4,42.1);
\draw[draw=none,fill=darkslateblue6460123] (axis cs:3.6,42.1) rectangle (axis cs:4.4,63);
\draw[draw=none,fill=darkslategray463058] (axis cs:6.6,63) rectangle (axis cs:7.4,63.6);
\draw[draw=none,fill=mediumaquamarine139217178] (axis cs:7.6,63.6) rectangle (axis cs:8.4,100);
\addplot [semithick, black, opacity=0.5]
table {%
-0.84 12.8000001907349
8.84 12.8000001907349
};
\addplot [semithick, black, opacity=0.5]
table {%
-0.84 27.2000007629394
8.84 27.2000007629394
};
\addplot [semithick, black, opacity=0.5]
table {%
-0.84 29.3999996185303
8.84 29.3999996185303
};
\addplot [semithick, black, opacity=0.5]
table {%
-0.84 35.0999984741211
8.84 35.0999984741211
};
\addplot [semithick, black, opacity=0.5]
table {%
-0.84 35.5
8.84 35.5
};
\addplot [semithick, black, opacity=0.5]
table {%
-0.84 39.2000007629394
8.84 39.2000007629394
};
\addplot [semithick, black, opacity=0.5]
table {%
-0.84 41.9000015258789
8.84 41.9000015258789
};
\addplot [semithick, black, opacity=0.5]
table {%
-0.84 42.0999984741211
8.84 42.0999984741211
};
\addplot [semithick, black, opacity=0.5]
table {%
-0.84 63
8.84 63
};
\addplot [semithick, black, opacity=0.5]
table {%
-0.84 63.5999984741211
8.84 63.5999984741211
};
\draw (axis cs:0,100.5) node[
  scale=0.8,
  anchor=base,
  text=black,
  rotate=0.0
]{12.6\%};
\draw (axis cs:1,100.5) node[
  scale=0.8,
  anchor=base,
  text=black,
  rotate=0.0
]{14.5\%};
\draw (axis cs:2,100.5) node[
  scale=0.8,
  anchor=base,
  text=black,
  rotate=0.0
]{2.2\%};
\draw (axis cs:3,100.5) node[
  scale=0.8,
  anchor=base,
  text=black,
  rotate=0.0
]{5.7\%};
\draw (axis cs:4,100.5) node[
  scale=0.8,
  anchor=base,
  text=black,
  rotate=0.0
]{27.7\%};
\draw (axis cs:5,100.5) node[
  scale=0.8,
  anchor=base,
  text=black,
  rotate=0.0
]{4.3\%};
\draw (axis cs:6,100.5) node[
  scale=0.8,
  anchor=base,
  text=black,
  rotate=0.0
]{3.3\%};
\draw (axis cs:7,100.5) node[
  scale=0.8,
  anchor=base,
  text=black,
  rotate=0.0
]{0.6\%};
\draw (axis cs:8,100.5) node[
  scale=0.8,
  anchor=base,
  text=black,
  rotate=0.0
]{36.4\%};
\end{axis}

\end{tikzpicture}
    \caption{Remaining crash causes for AEB for 1R~/~1V sensor set and unknown friction. All percentages are calculated based on the 925 simulated PCM cases, i.e. 100 \% is equivalent to 925 PCM cases. Det: Detection, Ego a.: ego acceleration, Fri: Friction, Ste: Steering, Opp a.: Opponent acceleration, N.c.: not classified, Avoided: avoided crashes.}
    \label{fig:AEB_crash_cause}
\end{figure}
In Figure~\ref{fig:AEB_crash_cause}, the crash causes for the AEB are depicted for the 1R~/~1V sensor set. 
As it can be observed, the crash cause with the highest percentage is Friction. Due to a reduced and unknown friction coefficient $\mu$, the assumed brake force of the AEB cannot be fully applied in many scenarios, creating a difference between the estimated braking distance and the actual braking distance. Therefore, the AEB is in 27.7~\% of cases not triggered in time to avoid the crash.
The second highest crash reason is due to a delay of the brake activation by the TTE condition with 14.5~\%. 
Late brake activation due to late Detection of the opponent plays with 12.6~\% another major role in the analyzed crash occurrences. In these cases, the crashes cannot be avoided because the opponent was not detected in time.
Minor roles play the TTC and the Ego acceleration triggering criteria as well as Steering and Opponent acceleration.

\subsubsection{V2X partial brake}\label{subsec:v2x_partial_brake}
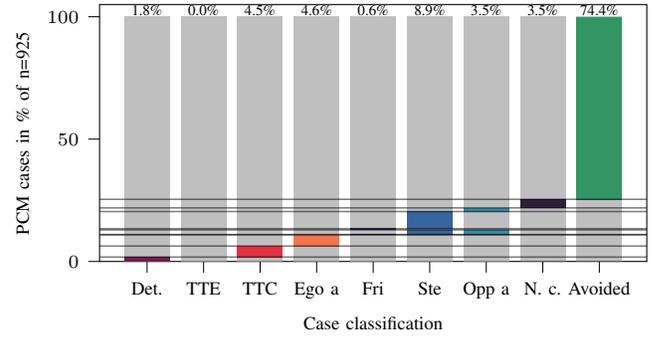
\begin{figure}[h]
    \centering
\begin{tikzpicture}

\definecolor{coral24311881}{RGB}{243,118,81}
\definecolor{crimson2255066}{RGB}{225,50,66}
\definecolor{darkgray176}{RGB}{176,176,176}
\definecolor{darkslateblue6460123}{RGB}{64,60,123}
\definecolor{darkslategray463058}{RGB}{46,30,58}
\definecolor{gray}{RGB}{128,128,128}
\definecolor{gray127}{RGB}{127,127,127}
\definecolor{mediumaquamarine139217178}{RGB}{50,150,100}
\definecolor{purple1113187}{RGB}{111,31,87}
\definecolor{steelblue51142167}{RGB}{51,142,167}
\definecolor{steelblue55101157}{RGB}{55,101,157}

\begin{axis}[
tick align=outside,
tick pos=left,
title={},
x grid style={darkgray176},
xlabel={Case classification},
xmin=-0.84, xmax=8.84,
xtick style={color=black},
xtick={0,1,2,3,4,5,6,7,8},
xticklabels={
  Det.,
  TTE,
  TTC,
  Ego a,
  Fri,
  Ste,
  Opp a,
  N. c.,
  Avoided
},
y grid style={darkgray176},
ylabel={PCM cases in \% of n=925},
ymin=0, ymax=105,
ytick style={color=black},
width=\linewidth,
height=5cm,
font=\scriptsize
]
\draw[draw=none,fill=gray127] (axis cs:-0.4,0) rectangle (axis cs:0.4,0);
\draw[draw=none,fill=gray127] (axis cs:0.6,0) rectangle (axis cs:1.4,0);
\draw[draw=none,fill=gray127] (axis cs:1.6,0) rectangle (axis cs:2.4,0);
\draw[draw=none,fill=gray127] (axis cs:2.6,0) rectangle (axis cs:3.4,0);
\draw[draw=none,fill=gray127] (axis cs:3.6,0) rectangle (axis cs:4.4,0);
\draw[draw=none,fill=gray127] (axis cs:4.6,0) rectangle (axis cs:5.4,0);
\draw[draw=none,fill=gray127] (axis cs:5.6,0) rectangle (axis cs:6.4,0);
\draw[draw=none,fill=gray,fill opacity=0.5] (axis cs:-0.4,0) rectangle (axis cs:0.4,100);
\draw[draw=none,fill=gray,fill opacity=0.5] (axis cs:0.6,0) rectangle (axis cs:1.4,100);
\draw[draw=none,fill=gray,fill opacity=0.5] (axis cs:1.6,0) rectangle (axis cs:2.4,100);
\draw[draw=none,fill=gray,fill opacity=0.5] (axis cs:2.6,0) rectangle (axis cs:3.4,100);
\draw[draw=none,fill=gray,fill opacity=0.5] (axis cs:3.6,0) rectangle (axis cs:4.4,100);
\draw[draw=none,fill=gray,fill opacity=0.5] (axis cs:4.6,0) rectangle (axis cs:5.4,100);
\draw[draw=none,fill=gray,fill opacity=0.5] (axis cs:5.6,0) rectangle (axis cs:6.4,100);
\draw[draw=none,fill=gray,fill opacity=0.5] (axis cs:6.6,0) rectangle (axis cs:7.4,100);
\draw[draw=none,fill=gray,fill opacity=0.5] (axis cs:7.6,0) rectangle (axis cs:8.4,100);
\draw[draw=none,fill=purple1113187] (axis cs:-0.4,0) rectangle (axis cs:0.4,1.8);
\draw[draw=none,fill=crimson2255066] (axis cs:1.6,1.8) rectangle (axis cs:2.4,6.3);
\draw[draw=none,fill=coral24311881] (axis cs:2.6,6.3) rectangle (axis cs:3.4,10.9);
\draw[draw=none,fill=darkslateblue6460123] (axis cs:3.6,10.9) rectangle (axis cs:4.4,11);
\draw[draw=none,fill=steelblue51142167] (axis cs:5.6,10.9) rectangle (axis cs:6.4,11);
\draw[draw=none,fill=steelblue55101157] (axis cs:4.6,11) rectangle (axis cs:5.4,12.9);
\draw[draw=none,fill=steelblue51142167] (axis cs:5.6,11) rectangle (axis cs:6.4,12.9);
\draw[draw=none,fill=darkslateblue6460123] (axis cs:3.6,12.9) rectangle (axis cs:4.4,13.4);
\draw[draw=none,fill=steelblue55101157] (axis cs:4.6,13.4) rectangle (axis cs:5.4,20.4);
\draw[draw=none,fill=steelblue51142167] (axis cs:5.6,20.4) rectangle (axis cs:6.4,21.9);
\draw[draw=none,fill=darkslategray463058] (axis cs:6.6,21.9) rectangle (axis cs:7.4,25.4);
\draw[draw=none,fill=mediumaquamarine139217178] (axis cs:7.6,25.4) rectangle (axis cs:8.4,99.8);
\addplot [semithick, black, opacity=0.5]
table {%
-0.84 1.79999995231628
8.84 1.79999995231628
};
\addplot [semithick, black, opacity=0.5]
table {%
-0.84 6.30000019073486
8.84 6.30000019073486
};
\addplot [semithick, black, opacity=0.5]
table {%
-0.84 10.8999996185303
8.84 10.8999996185303
};
\addplot [semithick, black, opacity=0.5]
table {%
-0.84 11
8.84 11
};
\addplot [semithick, black, opacity=0.5]
table {%
-0.84 12.8999996185303
8.84 12.8999996185303
};
\addplot [semithick, black, opacity=0.5]
table {%
-0.84 13.3999996185303
8.84 13.3999996185303
};
\addplot [semithick, black, opacity=0.5]
table {%
-0.84 20.3999996185303
8.84 20.3999996185303
};
\addplot [semithick, black, opacity=0.5]
table {%
-0.84 21.8999996185303
8.84 21.8999996185303
};
\addplot [semithick, black, opacity=0.5]
table {%
-0.84 25.3999996185303
8.84 25.3999996185303
};
\draw (axis cs:0,100.5) node[
  scale=0.8,
  anchor=base,
  text=black,
  rotate=0.0
]{1.8\%};
\draw (axis cs:1,100.5) node[
  scale=0.8,
  anchor=base,
  text=black,
  rotate=0.0
]{0.0\%};
\draw (axis cs:2,100.5) node[
  scale=0.8,
  anchor=base,
  text=black,
  rotate=0.0
]{4.5\%};
\draw (axis cs:3,100.5) node[
  scale=0.8,
  anchor=base,
  text=black,
  rotate=0.0
]{4.6\%};
\draw (axis cs:4,100.5) node[
  scale=0.8,
  anchor=base,
  text=black,
  rotate=0.0
]{0.6\%};
\draw (axis cs:5,100.5) node[
  scale=0.8,
  anchor=base,
  text=black,
  rotate=0.0
]{8.9\%};
\draw (axis cs:6,100.5) node[
  scale=0.8,
  anchor=base,
  text=black,
  rotate=0.0
]{3.5\%};
\draw (axis cs:7,100.5) node[
  scale=0.8,
  anchor=base,
  text=black,
  rotate=0.0
]{3.5\%};
\draw (axis cs:8,100.5) node[
  scale=0.8,
  anchor=base,
  text=black,
  rotate=0.0
]{74.4\%};
\end{axis}

\end{tikzpicture}
    \caption{Remaining crash causes for V2X partial brake and 2~s TTC threshold. All percentages are calculated based on the 925 simulated PCM cases, i.e. 100~\% is equivalent to 925 PCM cases. Det: Detection, Ego a.: ego acceleration, Fri: Friction, Ste: Steering, Opp a.: Opponent acceleration, n.c.: not classified, Avoided: avoided crashes.}
    \label{fig:v2x_brake_crash_causes}
\end{figure}
In Figure~\ref{fig:v2x_brake_crash_causes}, the crash reasons as well as the crash avoidance performance of the sole application of the V2X partial brake under a 2~s TTC threshold are depicted. As noted during the discussion of the overall results in Subsection~\ref{subsec:overall_crash_avoidance}, the V2X brake exhibits a higher crash avoidance performance than the AEB: 38.0~\% of crashes can be additionally avoided. The dominant crash reason for the V2X partial brake at the 2~s TTC threshold is Steering. This is explainable by considering the earlier point in time at which the brake needs to be triggered due to the reduced braking force of 4~\mpssq, especially in comparison to the full force AEB. Because of this longer time span between the crash-avoiding brake activation time point and the crash time point, the situation can change more drastically. Therefore, situations in which, for example, the opponent steers into the ego vehicle’s path at a point in time at which the brake cannot avoid the crash anymore, are more common for the partial brake than for the AEB. Second highest crash cause is Ego Acceleration, followed directly by the TTC threshold cause, while the Detection of the opponent due to the limited communication range plays a subordinate role. Note that for the V2X brake, the TTE condition was not considered in brake activation.

In Table~\ref{tab:V2X_brake_TTC_comparison}, the evaluated crash causes are depicted for simulations of the V2X brake simulation under varying TTC threshold values.
Comparing the 2~s TTC threshold results to the simulations under a 1.5~s TTC threshold, it can be observed that the crash avoidance performance decreases by 17.3~\% to 57.1~\%, which is, however, still significantly higher than the AEB performance of 36.4~\%, see Figure~\ref{fig:AEB_crash_cause}. The dominant crash cause for a 1.5~s TTC threshold is the TTC condition, accounting for 30.8~\%, exhibiting an increase of 26.3~\% compared to the 2~s TTC threshold. This result is to be expected when reducing the allowed time window before the crash by 0.5~s. All other crash causes exhibit a slight decrease.
By further decreasing the TTC threshold to 1.25~s before the crash, the crash avoidance performance decreases further to a value of 41.0~\%. As expected, the percentage of cases, in which the TTC restriction causes the crash, increases further to a value of now 50.9~\%.

Overall, the simulation results quantify the dependence of the V2X partial brake on the TTC threshold.  While the performances for thresholds of 2.0~s and 1.5~s still provide an improvement over the AEB results, limiting the TTC threshold to the same as the one of the AEB makes the crash avoidance performances of the two brake types comparable. 
\begin{table}[]
    \centering
    \caption{Remaining crash causes for V2X partial brake under different TTC thresholds. Results are represented in percent of the 925 simulated PCM cases. Det: Detection, Ego a.: ego acceleration, Fri: Friction, Ste: Steering, Opp a: Opponent acceleration, n.c.: not classified, Avoided: avoided crashes}
    \begin{tabular}{c|c c  c  c  c c  c  c  c}
                & Det   & TTE    & TTC    & Ego a & Fri   & Ste & Opp a      & n.c.   & avoided \\ \hline 
     2.0 s      & 1.8\% & 0.0\%  & 4.5\%  & 4.6\% & 0.6\% & 8.9\%    & 3.5\% & 3.5\% & 74.4\% \\
     1.5 s      & 0.2\% & 0.0\%  & 30.8\% & 4.6\% & 0.3\% & 3.8\%    & 1.2\% &  2.3\% & 57.1\%\\
     1.25 s     & 0.0\% & 0.0\%  & 50.9\% & 4.6\% & 0.1\% & 2.1\%    & 0.0\% & 1.3\%  & 41.0\%\\
    \end{tabular}
    
    \label{tab:V2X_brake_TTC_comparison}
\end{table}

\subsubsection{2-stage brake}\label{subsec:2-stage_brake}

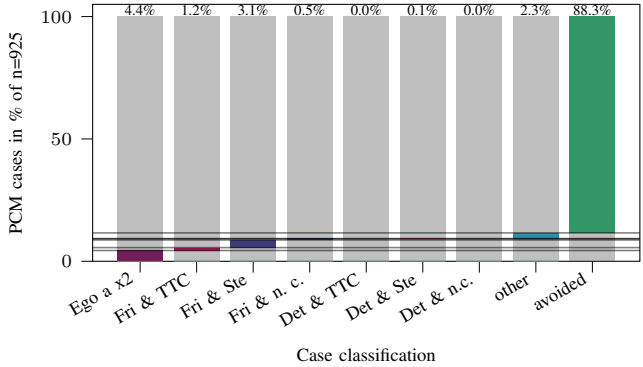
\begin{figure}[h]
    \centering
\begin{tikzpicture}

\definecolor{brown1722389}{RGB}{172,23,89}
\definecolor{coral24311881}{RGB}{243,118,81}
\definecolor{darkgray176}{RGB}{176,176,176}
\definecolor{darkslateblue6460123}{RGB}{64,60,123}
\definecolor{gray}{RGB}{128,128,128}
\definecolor{mediumaquamarine139217178}{RGB}{50,150,100}
\definecolor{purple1113187}{RGB}{111,31,87}
\definecolor{steelblue51142167}{RGB}{51,142,167}
\definecolor{steelblue55101157}{RGB}{55,101,157}

\begin{axis}[
tick align=outside,
tick pos=left,
title={},
x grid style={darkgray176},
xlabel={Case classification},
xmin=-0.84, xmax=8.84,
xtick style={color=black},
xtick={0,1,2,3,4,5,6,7,8},
xticklabel style={rotate=30.0,anchor=east},
xticklabels={
  Ego a x2,
  Fri \& TTC,
  Fri \& Ste,
  Fri \& n. c.,
  Det \& TTC,
  Det \& Ste,
  Det \& n.c.,
  other,
  avoided
},
y grid style={darkgray176},
ylabel={PCM cases in \% of n=925},
ymin=0, ymax=105,
ytick style={color=black},
width=\linewidth,
height=5cm,
font=\scriptsize
]
\draw[draw=none,fill=steelblue51142167] (axis cs:-0.4,0) rectangle (axis cs:0.4,0);
\draw[draw=none,fill=steelblue51142167] (axis cs:0.6,0) rectangle (axis cs:1.4,0);
\draw[draw=none,fill=steelblue51142167] (axis cs:1.6,0) rectangle (axis cs:2.4,0);
\draw[draw=none,fill=steelblue51142167] (axis cs:2.6,0) rectangle (axis cs:3.4,0);
\draw[draw=none,fill=steelblue51142167] (axis cs:3.6,0) rectangle (axis cs:4.4,0);
\draw[draw=none,fill=steelblue51142167] (axis cs:4.6,0) rectangle (axis cs:5.4,0);
\draw[draw=none,fill=steelblue51142167] (axis cs:5.6,0) rectangle (axis cs:6.4,0);
\draw[draw=none,fill=steelblue51142167] (axis cs:6.6,0) rectangle (axis cs:7.4,0);
\draw[draw=none,fill=gray,fill opacity=0.5] (axis cs:-0.4,0) rectangle (axis cs:0.4,100);
\draw[draw=none,fill=purple1113187] (axis cs:-0.4,0) rectangle (axis cs:0.4,4.4);
\draw[draw=none,fill=gray,fill opacity=0.5] (axis cs:0.6,0) rectangle (axis cs:1.4,100);
\draw[draw=none,fill=brown1722389] (axis cs:0.6,4.4) rectangle (axis cs:1.4,5.6);
\draw[draw=none,fill=gray,fill opacity=0.5] (axis cs:1.6,0) rectangle (axis cs:2.4,100);
\draw[draw=none,fill=darkslateblue6460123] (axis cs:1.6,5.6) rectangle (axis cs:2.4,8.7);
\draw[draw=none,fill=gray,fill opacity=0.5] (axis cs:2.6,0) rectangle (axis cs:3.4,100);
\draw[draw=none,fill=steelblue55101157] (axis cs:2.6,8.7) rectangle (axis cs:3.4,9.2);
\draw[draw=none,fill=gray,fill opacity=0.5] (axis cs:3.6,0) rectangle (axis cs:4.4,100);
\draw[draw=none,fill=gray,fill opacity=0.5] (axis cs:4.6,0) rectangle (axis cs:5.4,100);
\draw[draw=none,fill=coral24311881] (axis cs:4.6,9.2) rectangle (axis cs:5.4,9.3);
\draw[draw=none,fill=gray,fill opacity=0.5] (axis cs:5.6,0) rectangle (axis cs:6.4,100);
\draw[draw=none,fill=gray,fill opacity=0.5] (axis cs:6.6,0) rectangle (axis cs:7.4,100);
\draw[draw=none,fill=steelblue51142167] (axis cs:6.6,9.3) rectangle (axis cs:7.4,11.6);
\draw[draw=none,fill=gray,fill opacity=0.5] (axis cs:7.6,0) rectangle (axis cs:8.4,100);
\draw[draw=none,fill=mediumaquamarine139217178] (axis cs:7.6,11.6) rectangle (axis cs:8.4,99.9);
\addplot [semithick, black, opacity=0.5]
table {%
-0.84 4.40000009536744
8.84 4.40000009536744
};
\addplot [semithick, black, opacity=0.5]
table {%
-0.84 5.59999990463258
8.84 5.59999990463258
};
\addplot [semithick, black, opacity=0.5]
table {%
-0.84 8.69999980926514
8.84 8.69999980926514
};
\addplot [semithick, black, opacity=0.5]
table {%
-0.84 9.19999980926514
8.84 9.19999980926514
};
\addplot [semithick, black, opacity=0.5]
table {%
-0.84 9.30000019073487
8.84 9.30000019073487
};
\addplot [semithick, black, opacity=0.5]
table {%
-0.84 11.6000003814697
8.84 11.6000003814697
};
\draw (axis cs:0,100.5) node[
  scale=0.8,
  anchor=base,
  text=black,
  rotate=0.0
]{4.4\%};
\draw (axis cs:1,100.5) node[
  scale=0.8,
  anchor=base,
  text=black,
  rotate=0.0
]{1.2\%};
\draw (axis cs:2,100.5) node[
  scale=0.8,
  anchor=base,
  text=black,
  rotate=0.0
]{3.1\%};
\draw (axis cs:3,100.5) node[
  scale=0.8,
  anchor=base,
  text=black,
  rotate=0.0
]{0.5\%};
\draw (axis cs:4,100.5) node[
  scale=0.8,
  anchor=base,
  text=black,
  rotate=0.0
]{0.0\%};
\draw (axis cs:5,100.5) node[
  scale=0.8,
  anchor=base,
  text=black,
  rotate=0.0
]{0.1\%};
\draw (axis cs:6,100.5) node[
  scale=0.8,
  anchor=base,
  text=black,
  rotate=0.0
]{0.0\%};
\draw (axis cs:7,100.5) node[
  scale=0.8,
  anchor=base,
  text=black,
  rotate=0.0
]{2.3\%};
\draw (axis cs:8,100.5) node[
  scale=0.8,
  anchor=base,
  text=black,
  rotate=0.0
]{88.3\%};
\end{axis}

\end{tikzpicture}
    \caption{Remaining crash causes for 2 stage braking system with 1R~/~1V sensor set and 2s TTC threshold. All percentages are calculated based on the 925 simulated PCM cases, i.e. 100 \% is equivalent to 925 PCM cases. Det: Detection, Ego a: ego acceleration, Fri: Friction, Ste: Steering, Opp a: Opponent acceleration, N.c.: not classified, Avoided: avoided crashes.}
    \label{fig:2-stage_brake_crash_causes}
\end{figure}
In Figure~\ref{fig:2-stage_brake_crash_causes}, the results regarding the crash causes for the 2-stage braking system using the 1R~/~1V sensor set and a 2~s TTC threshold are displayed. As the 2-stage brake is a braking cascade consisting of the V2X brake and the AEB, one cause can be identified for each brake type. For the depiction, the crash causes of the two brake types are clustered. The first element in the description on the abscissa belongs to the AEB, while the second is associated with the V2X brake. For example, the second bar of Figure~\ref{fig:2-stage_brake_crash_causes} displays the percentage over cases where the crash cause for the AEB is Friction, while the crash cause of the V2X brake is the TTC threshold.\\
Because of the high crash avoidance performance of the 2-stage brake with 88.3~\% of avoided crashes, the percentages for the crash causes in the remaining crashes are low. Highest among them is ego acceleration for the two brake types. This cause is independent of braking force, opponent detection, sensor set and TTC threshold: If the ego vehicle is accelerating, the brake is not activated. Therefore, even the combination of the braking systems cannot avoid the crash in these cases.\\
The second largest group is the combination Friction \& Steering, making up a total 3.1~\%. By analyzing the separate results for AEB and V2X brake in Figure~\ref{fig:AEB_crash_cause} and Figure~\ref{fig:v2x_brake_crash_causes}, it can be observed that Friction was the main crash cause for the execution of the AEB, while steering was the main cause for the V2X partial brake under a 2~s TTC threshold. By combination of the two brake types, the crashes that are caused by Friction \& Steering are drastically reduced compared to the separate execution of the brakes.
\begin{table}[]
    \centering
    \caption{Remaining crash causes for 2-stage brake under different TTC thresholds and the 1R / 1V sensor set. Results are represented in percent of the 925 simulated PCM cases. Ego a.: ego acceleration, Fri: Friction, Ste: Steering,  avoided: avoided crashes}
    \begin{tabular}{c|c c  c  c  c }
                & Ego a x2   & Fri \& TTC    & Fri \& Ste     & other  & avoided  \\ \hline 
     2.0 s      & 4.4\%      & 1.2\%         & 3.1\%          & 2.9\% & 88.3\%  \\
     1.5 s      & 4.4\%      & 10.1\%        & 1.4\%          & 2.1\% & 81.9\% \\
     1.25 s     & 4.4\%      & 16.4\%        & 1.0\%          & 2.8\% & 75.5\% \\
    \end{tabular}
    
    \label{tab:2-stage_brake_comparison_TTC_thresholds}
\end{table}

For the analysis of the crash causes under different parameter variations, we focus on the dependence on the V2X TTC threshold due to the higher sensitivity of the brake under test regarding this parameter compared to using different classes of sensor sets, see Table~\ref{tab:overall_results}. The results are depicted in Table~\ref{tab:2-stage_brake_comparison_TTC_thresholds}. It can be observed that, by lowering the TTC threshold, the crash reason  Friction \& TTC  rises, while the remaining crash causes remain stable. However, the drop in crash avoidance performance of roughly 6.4~\% between the varied TTC thresholds is not as significant as it was the case for the V2X brake alone with an average difference of 16.7~\%, see Table~\ref{tab:V2X_brake_TTC_comparison}. Nevertheless, the TTC threshold and therefore, the application time window of the V2X brake is the most decisive parameter regarding the crash avoidance performance, both for the V2X partial braking system as well as for the 2-stage braking system.

\subsubsection{Assuming known friction coefficient}\label{subsec:assuming_friction_is_known}

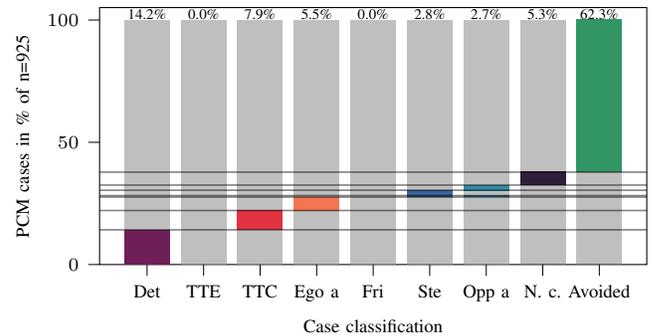
\begin{figure}[h]
    \centering
\begin{tikzpicture}

\definecolor{coral24311881}{RGB}{243,118,81}
\definecolor{crimson2255066}{RGB}{225,50,66}
\definecolor{darkgray176}{RGB}{176,176,176}
\definecolor{darkslategray463058}{RGB}{46,30,58}
\definecolor{gray}{RGB}{128,128,128}
\definecolor{gray127}{RGB}{127,127,127}
\definecolor{mediumaquamarine139217178}{RGB}{50,150,100}
\definecolor{purple1113187}{RGB}{111,31,87}
\definecolor{steelblue51142167}{RGB}{51,142,167}
\definecolor{steelblue55101157}{RGB}{55,101,157}

\begin{axis}[
tick align=outside,
tick pos=left,
title={},
x grid style={darkgray176},
xlabel={Case classification},
xmin=-0.84, xmax=8.84,
xtick style={color=black},
xtick={0,1,2,3,4,5,6,7,8},
xticklabels={
  Det,
  TTE,
  TTC,
  Ego a,
  Fri,
  Ste,
  Opp a,
  N. c.,
  Avoided
},
y grid style={darkgray176},
ylabel={PCM cases in \% of n=925},
ymin=0, ymax=105.105,
ytick style={color=black},
width=\linewidth,
height=5cm,
font=\scriptsize
]
\draw[draw=none,fill=gray127] (axis cs:-0.4,0) rectangle (axis cs:0.4,0);
\draw[draw=none,fill=gray127] (axis cs:0.6,0) rectangle (axis cs:1.4,0);
\draw[draw=none,fill=gray127] (axis cs:1.6,0) rectangle (axis cs:2.4,0);
\draw[draw=none,fill=gray127] (axis cs:2.6,0) rectangle (axis cs:3.4,0);
\draw[draw=none,fill=gray127] (axis cs:3.6,0) rectangle (axis cs:4.4,0);
\draw[draw=none,fill=gray127] (axis cs:4.6,0) rectangle (axis cs:5.4,0);
\draw[draw=none,fill=gray127] (axis cs:5.6,0) rectangle (axis cs:6.4,0);
\draw[draw=none,fill=gray,fill opacity=0.5] (axis cs:-0.4,0) rectangle (axis cs:0.4,100);
\draw[draw=none,fill=gray,fill opacity=0.5] (axis cs:0.6,0) rectangle (axis cs:1.4,100);
\draw[draw=none,fill=gray,fill opacity=0.5] (axis cs:1.6,0) rectangle (axis cs:2.4,100);
\draw[draw=none,fill=gray,fill opacity=0.5] (axis cs:2.6,0) rectangle (axis cs:3.4,100);
\draw[draw=none,fill=gray,fill opacity=0.5] (axis cs:3.6,0) rectangle (axis cs:4.4,100);
\draw[draw=none,fill=gray,fill opacity=0.5] (axis cs:4.6,0) rectangle (axis cs:5.4,100);
\draw[draw=none,fill=gray,fill opacity=0.5] (axis cs:5.6,0) rectangle (axis cs:6.4,100);
\draw[draw=none,fill=gray,fill opacity=0.5] (axis cs:6.6,0) rectangle (axis cs:7.4,100);
\draw[draw=none,fill=gray,fill opacity=0.5] (axis cs:7.6,0) rectangle (axis cs:8.4,100);
\draw[draw=none,fill=purple1113187] (axis cs:-0.4,0) rectangle (axis cs:0.4,14.2);
\draw[draw=none,fill=crimson2255066] (axis cs:1.6,14.2) rectangle (axis cs:2.4,22.1);
\draw[draw=none,fill=coral24311881] (axis cs:2.6,22.1) rectangle (axis cs:3.4,27.6);
\draw[draw=none,fill=steelblue55101157] (axis cs:4.6,27.6) rectangle (axis cs:5.4,28.2);
\draw[draw=none,fill=steelblue51142167] (axis cs:5.6,27.6) rectangle (axis cs:6.4,28.2);
\draw[draw=none,fill=steelblue55101157] (axis cs:4.6,28.2) rectangle (axis cs:5.4,30.4);
\draw[draw=none,fill=steelblue51142167] (axis cs:5.6,30.4) rectangle (axis cs:6.4,32.5);
\draw[draw=none,fill=darkslategray463058] (axis cs:6.6,32.5) rectangle (axis cs:7.4,37.8);
\draw[draw=none,fill=mediumaquamarine139217178] (axis cs:7.6,37.8) rectangle (axis cs:8.4,100.1);
\addplot [semithick, black, opacity=0.5]
table {%
-0.84 14.1999998092651
8.84 14.1999998092651
};
\addplot [semithick, black, opacity=0.5]
table {%
-0.84 22.1000003814697
8.84 22.1000003814697
};
\addplot [semithick, black, opacity=0.5]
table {%
-0.84 27.6000003814697
8.84 27.6000003814697
};
\addplot [semithick, black, opacity=0.5]
table {%
-0.84 28.2000007629395
8.84 28.2000007629395
};
\addplot [semithick, black, opacity=0.5]
table {%
-0.84 30.3999996185303
8.84 30.3999996185303
};
\addplot [semithick, black, opacity=0.5]
table {%
-0.84 32.5
8.84 32.5
};
\addplot [semithick, black, opacity=0.5]
table {%
-0.84 37.7999992370606
8.84 37.7999992370606
};
\draw (axis cs:0,100.5) node[
  scale=0.8,
  anchor=base,
  text=black,
  rotate=0.0
]{14.2\%};
\draw (axis cs:1,100.5) node[
  scale=0.8,
  anchor=base,
  text=black,
  rotate=0.0
]{0.0\%};
\draw (axis cs:2,100.5) node[
  scale=0.8,
  anchor=base,
  text=black,
  rotate=0.0
]{7.9\%};
\draw (axis cs:3,100.5) node[
  scale=0.8,
  anchor=base,
  text=black,
  rotate=0.0
]{5.5\%};
\draw (axis cs:4,100.5) node[
  scale=0.8,
  anchor=base,
  text=black,
  rotate=0.0
]{0.0\%};
\draw (axis cs:5,100.5) node[
  scale=0.8,
  anchor=base,
  text=black,
  rotate=0.0
]{2.8\%};
\draw (axis cs:6,100.5) node[
  scale=0.8,
  anchor=base,
  text=black,
  rotate=0.0
]{2.7\%};
\draw (axis cs:7,100.5) node[
  scale=0.8,
  anchor=base,
  text=black,
  rotate=0.0
]{5.3\%};
\draw (axis cs:8,100.5) node[
  scale=0.8,
  anchor=base,
  text=black,
  rotate=0.0
]{62.3\%};
\end{axis}

\end{tikzpicture}
    \caption{Remaining crash causes for AEB for 1R~/~1V sensor set. Friction coefficient is assumed to be known and TTE is removed from brake trigger conditions. All percentages are calculated based on the 925 simulated PCM cases, i.e. 100 \% is equivalent to 925 PCM cases. Det: Detection, Ego a.: ego acceleration, Fri: Friction, Ste: Steering, Opp a: Opponent acceleration, n.c.: not classified, Avoided: avoided crashes.}
    \label{fig:AEB_known_mu_crash_cause}
\end{figure}

Friction and the TTE condition are the two main crash reasons for the AEB system as indicated in Figure~\ref{fig:AEB_crash_cause}. The many cases of unavoidable crashes due to friction stem from the fact that the road friction coefficient is often reduced, averaging at 0.81, with a minimum value at 0.18. Accordingly, the brake force is limited in many cases to values below the expected 9~\mpssq. The AEB system, however, plans the brake application assuming that the full force is applicable. Therefore, the actual braking distance is larger than the estimated distance, wherefore the crashes cannot be avoided.
It can be assumed that if the friction coefficient would be known by the system, the crash avoidance performance could be increased. To validate this hypothesis, the PCM cases are simulated under the assumption that the friction coefficient is known by the AEB system, enabling the system to correctly estimate the braking force that can be applied. In addition to that, we remove the TTE conditions from the brakes trigger list. The results for the 1R~/~1V sensor set are depicted in Figure~\ref{fig:AEB_known_mu_crash_cause}. The differences regarding the crashes between the two simulations are summarized in Table~\ref{tab:AEB_known_mu_comparision}.
\tabcolsep=0.08cm
\begin{table}[]
    \centering
    \caption{Difference in crash causes regarding simulation of AEB under 1R~/~1V sensor set with unknown friction/TTE condition (UF/TTE) and known friction/no TTE condition (KF). Note that the sum of the changes does not add to 100\% because the crash reasons can overlap for friction and secondary causes. All vales are in percent of the 925 simulated PCM cases.}
    \begin{tabular}{c|c c  c  c  c c  c  c  c}
                & Det   & TTE    & TTC      & Ego a  & Fri     & Ste      & Opp a      & n.c.   & avoided \\ \hline 
     UF/     & \multirow{2}{*}{12.6\%}  &  \multirow{2}{*}{14.5\%}   & \multirow{2}{*}{2.2\%}    & \multirow{2}{*}{5.7\% }   & \multirow{2}{*}{27.7\% }   &  \multirow{2}{*}{4.3\% }    & \multirow{2}{*}{3.3\%  }       & \multirow{2}{*}{0.6\%  }   & \multirow{2}{*}{36.4\% }  \\
     TTE        &&&&&&&&& \\ 
     KF         & 14.1\% & 0.0\%   & 7.9\%  & 5.5\%  & 0\%     & 2.9\%    & 2.8\%      &  5.2\% & 62.3\%\\

    \end{tabular}
    
    \label{tab:AEB_known_mu_comparision}
\end{table}
As expected, there are now no crashes that are caused by friction and the TTE condition. Under the described assumption, the percentage of avoided crashes rises from 36.4~\% to 62.3~\%, which is an increase of 25.9~\%. Additionally, the crash causes for the TTC condition as well as the Detection increases. This can be anticipated: If the brake force is reduced to some value below 9~\mpssq\ due to reduced friction, then the braking distance increases and therefore the brake needs to be activated earlier in time before the crash. In 5.7~\% of cases, this pushes the needed trigger time point beyond the TTC threshold of 1.25~s. And in 1.6~\% of cases, the opponent is still occluded by the view obstruction, when the triggering time is in the allowed time window set by the TTC threshold, creating a rise in the Detection crash cause. 

\subsection{Comparison to other studies}
In \cite{Scanlon2017}, a similar analysis is performed for an onboard sensor triggered AEB, while V2X-enhanced braking systems are not considered. 448 straight crossing path crashes from a US study are simulated, including cases with reduced road friction. The assumed sensor set corresponds to the 1R~/~1V sensor set of this paper's analysis. Under a TTC threshold of 1.0~s and a computational latency of 0.25~s, 31~\% of crashes are avoided in \cite{Scanlon2017}. When adjusting our simulation to a similar parameterization, the crash avoidance rate for the 925 PCM cases lies at 34.3~\% and is therefore comparable with the results of \cite{Scanlon2017}.

The works \cite{Khayyat2022} and \cite{Lai2023} evaluate a V2X-enhanced AEB system, which a achieve a 100~\% crash avoidance rate in a limited number of 90° straight crossing path scenarios. \cite{Avino2018} reports similar results: Assuming a 100~\% V2X penetration rate and a simple channel model, i.e. the most comparable case to our study, about 95~\% of potential collisions are detected in time to prevent the crash. Diverse driver behavior, e.g. steering (for \cite{Khayyat2022} \cite{Lai2023} \cite{Avino2018}) and acceleration (for \cite{Khayyat2022} \cite{Lai2023}), is not considered. In our earlier work  \cite{Zimmermann2024}, where these effects are neglected as well, the 2-stage brake achieves a 100~\% crash avoidance rate under a 2~s TTC threshold. However, the study of this paper shows that these real world factors, namely the crash reasons Ego Acceleration and Opponent Steering, have a negative influence on the crash avoidance performance, reducing the percentage of avoided crashes to a maximum of 88.3~\%.

\section{Conclusion}\label{sec:conclusion}
The use of a V2X-enhanced braking system, compared to a onboard sensor-triggered automatic emergency brake shows significant improvements in crash avoidance performance for the simulated test cases, which depict real-world crash situations from the GIDAS database. \\
Based on the results presented in this paper, an extrapolation to the overall crash avoidance potential in Germany per year can be made, using the associated case weights in the GIDAS data base. Additionally, other factors, such as data and prediction uncertainty, need to be included in further studies.

\bibliographystyle{IEEEtran} 
\bibliography{refs} 

\vspace{12pt}
\color{red}

\end{document}